\documentclass[useAMS,usenatbib]{mn2e}
\usepackage{amsmath}
\usepackage{times}
\usepackage{graphicx}
\usepackage{wasysym}

\title[Dark Matter Annihilation in the First Galaxies]{Dark Matter Annihilation in the First Galaxy Halos}

\author[Sch\"{o}n et. al.]{S. Sch\"{o}n$^{1}$\thanks{E-mail:
     sschon@student.unimelb.edu.au},  K. J. Mack$^{1, 2, 3}$, C. A. Avram$^{1, 3}$, J. S. B. Wyithe$^{1, 2}$ and E. Barberio$^{1, 3}$\\ \\ $^1$ School of Physics, The University of Melbourne, Parkville, VIC 3010, Australia \\ $^2$ ARC Centre of Excellence for All-Sky Astrophysics (CAASTRO) \\ $^3$ ARC Centre of Excellence for Particle Physics at the Terascale (CoEPP) \\}

\begin{document}

\maketitle

\begin{abstract}
We investigate the impact of energy released from self-annihilating dark matter on heating of gas in the small, high-redshift dark matter halos thought to host the first stars. A SUSY neutralino like particle is implemented as our dark matter candidate.  The PYTHIA code is used to model the final, stable particle distributions produced during the annihilation process. We use an analytic treatment in conjunction with the code MEDEA2 to find the energy transfer and subsequent partition into heating, ionising and Lyman alpha photon components. We consider a number of halo density models, dark matter particle masses and annihilation channels. We find that the injected energy from dark matter exceeds the binding energy of the gas within a $10^{5}$-$10^{6}\mathrm{M}_{\odot}$ halo at redshifts above 20, preventing star formation in early halos in which primordial gas would otherwise cool. Thus we find that DM annihilation could delay the formation of the first galaxies. 
\end{abstract}


\section{Introduction}
Dark matter is recognised as an integral part of modern concordance cosmology, and while an empirical characterisation has had considerable success in the description of wider astrophysical phenomena such as structure formation and gravitational lensing, dark matter's precise nature has remained unclear. Although there exists no suitable candidate within the existing standard model paradigm, an elementary particle formulation of dark matter remains the favoured solution (as opposed to a modified gravity model). As such, the dark matter problem is of current interest to both astro and particle physics, an overlap which may create unique opportunities in trying to establish the fundamental form of dark matter. In addition, models of physics beyond the standard model have proven fruitful sources of viable dark matter candidates, making dark matter a valuable road marker on the path to a complete theoretical formulation of fundamental physics. For general reviews of these topics see \citet{Roos2012}, \citet{Bertone2004} and \citet{Bergstrom2012}.

Our primary understanding of dark matter comes from the observation of astrophysical phenomena. A range of projects probing dark matter's behaviour beyond gravitational interactions are currently in progress or in the planning stages, including collider (\citealt{Birkedal2004}, \citealt{Collider2001}, \citealt{Allanach2000} and \citealt{Baer2003}) and direct detection experiments (\citealt{Aalseth2013}, \citealt{Ahmed2011}, \citealt{Angle2011} and \citealt{Armengaud2012}). While there are positive indicators that these will be able to constrain and confirm dark matter models in the future, data released to date has not proven conclusive [see for example, the comparison between DAMA/LIBRA \citep{Bernabei2010} and LUX \citep{LUX2014}].

Another avenue of investigation is that of indirect detection in which dark matter is assumed to produce standard model particles through non-gravitational interaction such as decay or annihilation. Particularly dense regions, such as are found at the centres of massive galaxies or clusters, could produce distinct gamma or x-ray signatures or signals from other particle excesses (\citealt{Grasso2009}, \citealt{Bergstrom2009} and \citealt{Bertone2006} and \citealt{Prada2004}). Unambiguous identification as products of dark matter annihilation is complicated by the presence of other astrophysical sources such as pulsars and supernova which may mimic such a signal (\citealt{Hooper2009} and \citealt{Biermann2009}). Alternatively one may consider a more global impact by examining how the extra energy from dark matter annihilation affects features such as the high redshift 21cm signal from the galactic medium (\citealt{Evoli2014}, \citealt{Sitwell2014}). Modifications may be particularly distinct during the early era of reionisation where the power from dark matter energy injection was not swamped by astrophysical energy sources. 

When introducing dark matter models into wider cosmological calculations, a number of complexities need to be taken into consideration. These include the inherent uncertainties in both the particle and astrophysics models \citep{Mack2014}, impact on standard astrophysical processes \citep{Fontanot2014} and possibly exotic structure formation \citep{Spolyar2008}. The impact dark matter annihilation has on star formation is addressed in (\citealt{Ripamonti2007}, \citealt{Natarajan2009}, \citealt{Ripamonti2010}, \citealt{Stacy2012} and \citealt{Smith2012}). To further this line of investigation, we look to incorporate the physical processes implemented in the updated MEDEA2 code to account for the injection of relativistic particles produced by massive DM particles. The energy transfer routine employed in this work allows us to explore the impact different dark matter halo models, as well as variation in the dark matter annihilation products, have on the total energy deposited over a wide range of redshifts and halo masses.

In this paper we investigate the energy transfer from dark matter self-annihilation in dark matter halos at high redshifts and the impact this may have on gas within the halo. This is of interest for several reasons. Firstly, the existence of a self-heating source could impact early star formation which in turn would have wider implications for the gas in the IGM and the rate of reionisation. Secondly, the contribution from small, collapsed structure to the overall energy produced by dark matter annihilation is considerable so careful treatment of the energy transfer is desirable.  

An outline of the paper is as follows. We begin with a summary of our method in \S 2. We describe a model of dark matter halos across a range of masses and redshifts in  \S 3. Descriptions of the simulated final stable states of the dark matter annihilation process, and the appropriate first order analytic approximation of the energy transfer between the injected particles and the halo's gas component are given in \S 4 and 5 respectively. We discuss the comparison between the injected energy and the halo's gravitational binding energy in \S 6, and the heating of the diffuse gas surrounding the halo in \S 7. We conclude with a discussion in \S 8.

\section{Method} 
We model the cosmological dark matter component using simple analytic expressions which allow for a straightforward exploration of the possible parameter space.  The uncertainties in these quantities are of particular interest as we wish to minimise the possibility of astrophysical sources creating results which are degenerate to variations of the dark matter model. 

Besides the halo model, the other key aspect in gauging the impact of self-annihilating DM on the halo's gas is a precise treatment of the injected energy produced by the DM annihilation process. This entails both the energy partition of the stable annihilation end products and the way these particles interact with the surrounding gas. We use PYTHIA (\citealt{PYTHIA2006}, \citealt{PYTHIA2008}) to simulate the self-annihilation of a SUSY-neutralino like particle and in this way also produce the spectral energy distribution of the injected particles. 

We approach the actual energy transfer calculation through path-averaged integrals of individual particles, assuming that the injected particles are oriented isotropically and travel in straight paths. (Our method does not account for change in path due to scattering events, thus underestimating the length of the total path spent within the halo.) A Monte Carlo method is used to sample the different paths the particle could have taken to reach the edge of the halo, and subsequently we arrived at the average energy lost by an electron, positron or photon of energy $E_{i}$, injected at radius $r_{i}$, once it reaches the virial radius. Our treatment of the relevant physical interactions follows \citealt{Evoli2012}, focusing on the primary injected particle. For a detailed discussion of the relevant processes see Appendix B. 

We then intergrate over the spatial and energy distributions of the particles to arrive at an estimate of the gross energy transferred to the gas. The Monte Carlo Energy DEposition (MEDEA) code (\citealt{Valdes2010}, \citealt{Evoli2012}), is then used to gauge how this energy is partitioned into heating, ionisation and Lyman alpha photons and how this could practically impact the halo's environment (see \S 6.1 for further details).

Throughout we take our cosmological parameters from Planck \citep{Planck} such that $h = 0.71$, $\Omega_{\Lambda,0} = 0.6825$ and $\Omega_{m,0} = 0.3175$.

\section{Dark Matter Halo Parameters}
While the density profiles of halos with mass upwards from $10^8 \mathrm{M}_{\odot}$ are relatively well explored in simulations (see for example \citealt{Merritt2005}, \citealt{Merritt2006}, \citealt{Zhao2009}, \citealt{Navarro2010} and \citealt{Salvador2012}), the precise form of smaller objects is less certain. Since these low-mass halos provide a significant boost to the overall injected energy from dark matter annihilation as well as playing host to early star formation, we consider both different density profiles and mass-concentration relations in their description. Our models of small halos are thus not necessarily definitive, physical representations but rather meant to cover a plausible parameter space. Halo masses under consideration range from $10^{3}\mathrm{M}_{\odot}$  - $10^{9}\mathrm{M}_{\odot}$ for redshift $0$-$50$.

\subsection{Halo Profiles}
We compare three different profiles. The NFW \citep{Navarro1995} (eq. 1) and Einasto \citep{Einasto1965} (eq. 2) profiles are qualitatively similar in so far they feature a density cusp at the centre of the halo, while the Burkert \citep{Burkert1995} (eq. 3) profile has a flattened core. As we found the NFW and Einasto profiles to show similar behaviour in later calculations, we shall only present the Einasto profile as the representative of cuspy halos. We further assume that halo density profiles remain self-similar across both mass and redshift range, neglecting halo assembly history. The profiles of NFW, Einasto and Burkert halos respectively are
\begin{eqnarray}
 \rho_{NFW}(r) = \frac{\rho_{0}}{(\frac{r}{r_{s}})(1+\frac{r}{r_{s}})^{2}}
\end{eqnarray}
\begin{eqnarray}
 \rho_{E} = \rho_{0}e^{-\frac{2}{\alpha_{e}}[(\frac{r}{r_{s}})^{\alpha}-1]}
\end{eqnarray}
\begin{eqnarray}
 \rho_{B} = \frac{\rho_{0}}{(1+\frac{r}{r_{s}})(1+\frac{r^{2}}{r_{s}^{2}})}
\end{eqnarray} 
In all cases $r_{s}$ is the scale radius and defined as $\frac{r_{vir}}{c}$, where $c$ is the concentration parameter (see  \S3.2). We here adopt the convention of defining the virial radius $r_{vir}$ of the halo as encompassing a spherical volume with density 200 times greater than the critical density. For the Einasto model, $\alpha_{e}=0.17$, and is here taken to be independent of mass. Lastly, $\rho_{0}$ is a normalisation constant such that the mass enclosed within the virial radius gives the total halo mass.

\subsection{Baryonic Profile}
We assume the gas component of the halo to follow the dark matter density distribution with a baryon fraction of $f_{b}=0.15$. In our treatment, we find that the energy deposited into the halo is predominantly driven by high energy electrons and positrons inverse Compton scattering off of CMB photons and the absorption of the secondary photons produced in this process. The CMB photon density is independent of the distribution of the baryonic matter and the secondary, up-scattered photons more readily interact even with lower density gas than the original, high energy particles injected. This reduces the impact the baryonic profile has on the total energy transferred. In this work, we find that the distribution of the dark matter, particularly at the centre of the halo, plays a more significant role than the baryons, in how much energy is deposited into the halo. For a comparison between this approach and a baryonic core \citep{Abel2002} profile see Appendix C.

\subsection{Mass-Concentration relations}
The concentration parameter sets the radius $r_{s}$ at which the density profiles turns over and as such regulates the density at the centre of the halo. We choose two contrasting, slightly modified expressions for the concentration-mass relation from \citet{Comerford2007} eq. 4 and  \citet{Duffy2008} eq. 5 which are both dependent on halo mass and redshift. The gradient of the relation from Comerford is considerably steeper than that of Duffy and both relations produce highly concentrated profiles for small mass halos at low redshift. Qualitatively, this behaviour persists to high redshifts, though the concentration parameter decreases overall (see Figure 1).  
\begin{eqnarray}
c_{c}(M,z) = \left( \frac{M}{1.3 \times 10^{13}M_{\odot}} \right)^{-0.15} \frac{22.5 }{(1+z)}
\end{eqnarray}
\begin{eqnarray}
c_{d}(M,z) = \left( \frac{M}{2 \times 10^{12}h^{-1}M_{\odot}} \right)^{-0.084}\frac{14.85}{(1+z)^{0.71}}
\end{eqnarray}
We note that both concentration relations were fitted for galaxy-sized halos at low redshift and we extrapolate considerably beyond their intended parameter space. As a check, we thus also consider a third, mass-independent modification of the above relations :
\begin{eqnarray}
c_{f}(z) = \frac{47.85}{(1+z)^{0.61}}.
\end{eqnarray}
In conjunction with the above density profiles these allow us to model a range of halo density distributions to investigate how halo morphology impacts dark matter annihilation effects.

\begin{figure}
\centering
\includegraphics[scale=0.45]{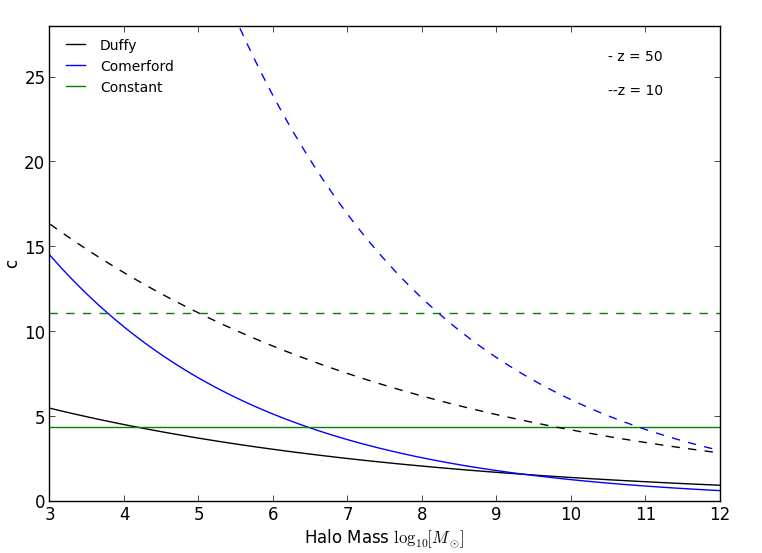}
\caption{Mass-concentration relation for halos between $10^{3}-10^{12}\mathrm{M}_{\odot}$ at redshift $50$ (solid line) and $10$ (dashed line). Black, blue and green refer to the Duffy, Comerford and constant models respectively.}
\label{Concentration}
\end{figure}

\section{Dark Matter Model}
We here choose a generic self-annihilating SUSY neutralino as our dark matter particle spanning masses from 5 GeV to 110 GeV, with quark/anti-quark pairs, muons, tau, or W bosons as their immediate annihilation products. In practice, more exotic candidates that either annihilate or decay to inject energy in the form of standard model particles could also be used provided a sufficient compatibility with the assumed cosmology.
\subsection{Dark Matter Annihilation Power}
The power produced by dark matter annihilation per unit volume is given by:
\begin{eqnarray}
P_{dm}(x) = \frac{c^{2}}{m_{dm}}  \langle v \sigma  \rangle \rho_{dm}^{2}(x).
\end{eqnarray}
where $m_{dm}$ and $\rho_{dm}$ are the dark matter particle mass and volume density respectively and $ \langle v \sigma \rangle $ is the velocity averaged annihilation cross-section which we take to be $2 \times 10^{-26} \mathrm{cm}^{-3}\mathrm{ s}^{-1} $. \footnote{While we here adopt a cross-section constant across all our models, we note that $ \langle v \sigma \rangle $ for some of our models may already be subject to constraints in conjunction with the dark matter particle mass employed.}
\subsection{Final Particle States}
PYTHIA is used to produce the final particle states for the various candidates. The dark matter annihilation event is simulated via an electron/positron proxy where the centre of mass energy of the collision is set as twice the mass of the dark matter particle. We further differentiate between annihilation paths via quark/anti-quark pairs, mu and tau leptons and W bosons. The annihilation products are simulated until only stable particles (neutrinos, electrons/positrons, photons, protons/ anti-protons) are left. Figure A1 shows the respective spectra of the energy distributions of electrons, positrons and photons for different dark matter masses and annihilation channels.

\section{Energy Transfer}
Energy is transferred to the halo's gas component via photons, electrons and positrons. Neutrinos only interact weakly and only negligible numbers of protons/anti-protons are created, so their contribution to the deposited energy is negligible.

\begin{figure*}
\centering
\includegraphics[scale=0.67]{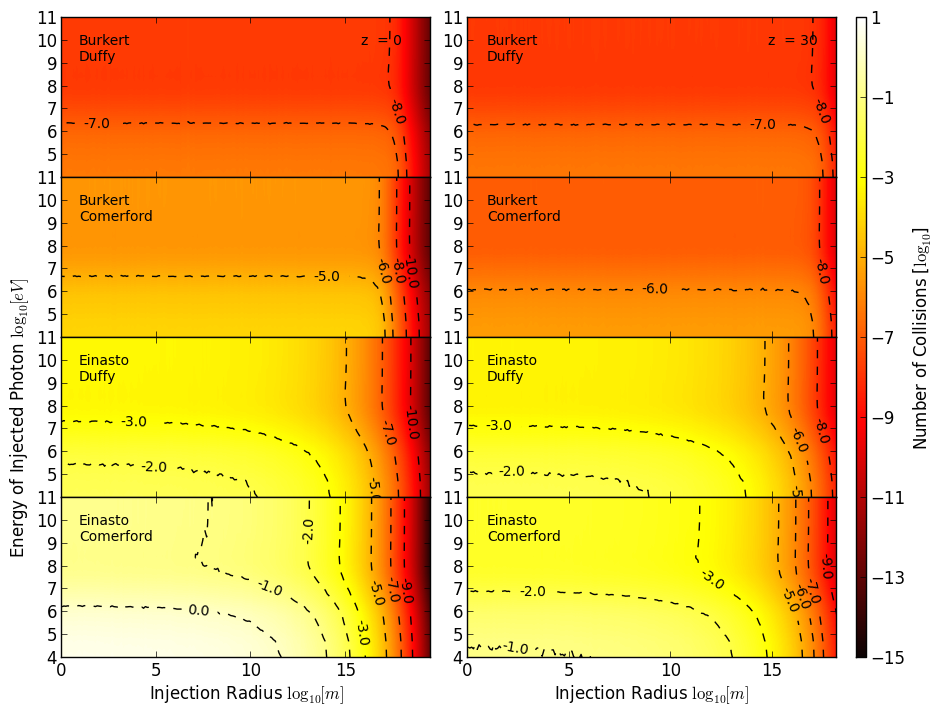}
\caption{Average number of collisions undergone by photons before reaching the virial radius in a $10^{5}\mathrm{M}_{\odot}$ halo. The left hand side shows halos at redshift 0 and the right at redshift 30. From upper to lower (least to most concentrated), the rows correspond respectively to a halo model of Burkert/Duffy, Burkert/Comerford, Einasto/Duffy and Einasto/Comerford. In each plot the contour lines correspond to the equivalent value given by the colourbar.}
\label{electron loss}
\end{figure*}

\subsection{Individual Particle Energy Loss}
For simplicity we assume that all particles retain a straight line trajectory as they travel through the halo.  While in reality the particles would undergo scattering processes, the approximation is justifiable as the gas in the halo is sparse enough so that only a small number of such scattering events occurs in the case of photons and the energy of the electrons and positrons undergoing inverse Compton (IC) scattering is much greater than that of the scattered cosmic microwave background (CMB) photons \citep{Evoli2012}.
The path of the particle is parametrised in the following way 
\begin{eqnarray}
\bmath{x}(t) = \bmath{x}_{f}t + \bmath{x}_{i}(1-t)
\end{eqnarray}
where $\bmath{x}_{i}$ and $\bmath{x}_{f}$ are its initial and final position, and $t \in [0,1]$.\\
The energy lost along the path is dependent on the type of particle, $\alpha$, its initial energy, $E_{in}$, and the density of the gas, $\rho_{g}$, it encounters and is given here by
\begin{eqnarray}
L_{\alpha}(E_{in}, \bmath{x}_{i}, \bmath{x}_{f}) = \int_{0}^{1}S_{\alpha}(E(t)) \rho_{g}(\bmath{x}(t)) \mathrm{d}t.
\end{eqnarray}
We make use of the spherical symmetry of the halo and calculate the average energy lost by a particle of species $\alpha$ created at radius $r_{i}$ (placed along the $x$-axis for convenience) while traveling to $r_{f}$

\begin{figure*}
\centering
\includegraphics[scale=0.67]{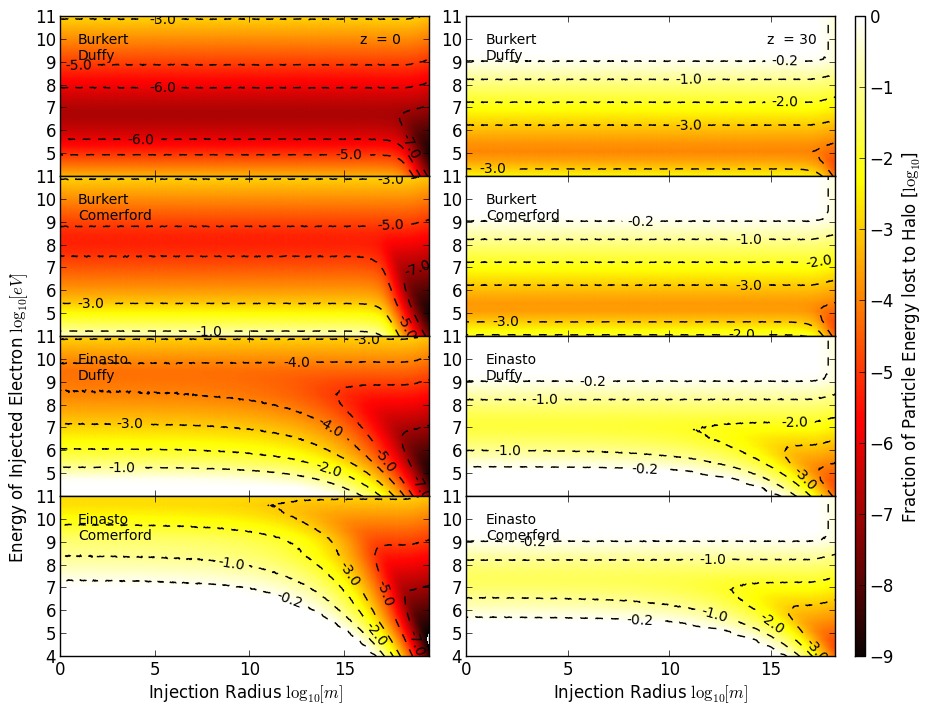}
\caption{Average fraction of energy lost by electron/positron created at radius $r$ before reaching the virial radius in a $10^{5}\mathrm{M}_{\odot}$ halo. The left hand side shows halos at redshift 0 and the right at redshift 30. From upper to lower (least to most concentrated), the rows correspond respectively to a halo model of Burkert/Duffy, Burkert/Comerford, Einasto/Duffy and Einasto/Comerford. The contour lines correspond to the equivalent value given by the colourbar.}
\label{electron loss}
\end{figure*} 

\begin{eqnarray}
\bar{L}_{\alpha}(E_{in}, r_{i}, r_{f}) = \frac{ \int_{0}^{\pi}  \int_{0}^{2 \pi}  L_{\alpha}(E_{in}, \bmath{x}(r_{i}) , \bmath{x}(r_{f}, \theta, \phi)) \mathrm{d}\theta \mathrm{d}\phi}{4 \pi r_{f}}
\end{eqnarray}

The energy loss rate for a photon is driven by the total interaction cross-section which is heavily dependent on the photon energy \citep{Review2012}. Figure 2 shows the number of interactions the photons with different energies, injected at various radii, undergo before escaping the halo. We note that in the case of high energy photons that predominantly lose energy through  electron/positron pair-creation,  particles will largely escape the halo without significant interaction. In contrast, for photons with energy below the $MeV$ range, the main energy transfer mechanisms moves to Compton scattering and photo-ionisation/excitation. These have a higher interaction cross-section than the pair-production process and so are considerably more efficient at depositing their energy into the gas. Overall only particles created very close to the core of the halo, and thus injected in a high density gas environment, contribute to the energy transfer in any notable form. Thus density profiles with a cusp and high mass-concentration parameters are considerably more efficient than the more relaxed models at depositing energy as they provide the high density core required for the photon energy-loss processes.

In contrast, electrons and positrons are assumed to lose energy continuously according to the particle's stopping power as well as in collisions via IC scattering off cosmic microwave background (CMB) photons. The latter process dominates in the high redshift regime. Figure 3 shows the fraction of the injected particle's initial energy that is lost as the particle reaches the virial radius, with the right hand side showing halos at redshift 30 and the left at redshift 0. From top to bottom the halo density profiles are ordered from least to most concentrated. We note that IC scattering is indeed shown to be the dominant energy-loss mechanism for high energy electrons at high redshift and the energy lost is independent of the halo profile. In contrast, energy loss through interactions with the halo gas is dominant for low-energy particles and is, as expected, more efficient in the highly concentrated models. While most of the injected electrons are high energy particles, and will therefore undergo IC scattering at high redshift, the lost energy will be transferred to the halo's gas through the up-scattered CMB photons created in the process. Thus while the energy loss of the injected electrons from IC scattering is independent of the halo profile, the amount of energy absorbed by the halo will still be dependent on its density distribution.

As a whole we find electrons and positrons to be the dominant source of dark matter annihilation energy being transferred to gas in the halo.

\begin{figure*}
\centering
\includegraphics[scale=0.7]{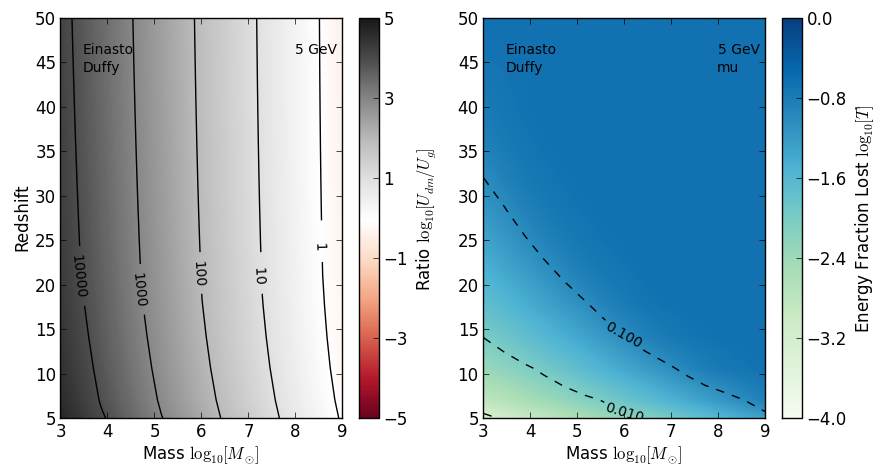}
\caption{The left panel shows the ratio of the total energy produced through dark matter annihilation and the gravitational binding energy of the gas. The right shows the fraction of energy lost by the particles as they leave the halo. Both use a halo model consisting of an Einasto profile with a Duffy mass concentration relation and a 5GeV dark matter particle. Again the contour lines correspond to the values given by the colourbar.}
\label{Electron Transfer}
\end{figure*}

\subsection{Total Energy Lost}
We now calculate the total fraction of the annihilation energy lost by the injected particles within the halo which requires integration over the energy spectrum at each point in the halo volume:

\begin{eqnarray}
E_{lost} = \sum_{\alpha=1}^{3} f_{\alpha} \int_{0}^{\infty} \int_{0}^{r_{vir}} 4 \pi \mu_{\alpha} \bar{L}_{\alpha}(\epsilon_{i}, r, r_{vir}) P_{dm}(r)r^{2}\mathrm{d}\epsilon \mathrm{d}r,
\end{eqnarray}
where $\alpha$ refers to the different injected particle species and $\mu_{\alpha}$ the fraction of the total annihilation energy in form of that species. Given that the total energy produced through dark matter annihilation is given by
\begin{eqnarray}
E_{tot} = \int_{0}^{r_{vir}}4 \pi P_{dm}(r)r^{2}\mathrm{d}r,
\end{eqnarray}
the fraction of the total energy from dark matter that is in turn lost in the halo is then simply
\begin{eqnarray}
T(M, z) = \frac{E_{lost}}{E_{tot}}.
\end{eqnarray}
It should be noted that high-energy particles create a cascade of lower-energy particles as they lose energy. Following these secondary particles is beyond the complexity of this calculation and we thus set an energy absorption fraction for these secondary particles (see \S 6 for further detail). These secondary particles carry less energy\footnote{The precise spectrum of secondary particles produced will be dependent on the down-scattering process.} than the originally injected particles and thus interact more readily with the gas than their high energy progenitors. 

\section{Binding Energy Comparison}
As an initial measure of the impact that dark matter annihilation has on the halo structure, we compare the total energy produced via annihilation to the halo's gravitational binding energy. 
The gravitational binding energy is given by
\begin{eqnarray}
U_{G} =  \int_{0}^{r_{vir}} \frac{G M_{int}(r')m_{shell}(r')}{r'} \mathrm{d}r',
\end{eqnarray}
and the total energy injected via dark matter annihilation over the Hubble time $t_{H}$ (which is here taken as a proxy for the halos age) is
\begin{eqnarray}
U_{dm} = \int_{0}^{r_{vir}}4\pi P_{dm}(r)r^{2}dr \bmath{\cdot} t_{H}.
\end{eqnarray}

In Figure 4 we plot the ratio between the annihilation energy and the binding energy (left panel) and the fraction of the annihilation energy lost to the halo (right panel) for an Einasto profile with a Duffy mass-concentration relation and a 5 GeV dark matter particle. We note that in small halos the annihilation energy is an order of magnitude larger than the binding energy when taken over the Hubble time (the increase in Hubble time also accounts for the ratio increasing at low redshift). In contrast the transfer of energy to the halo is more efficient in large halos at high redshift which is consistent with IC scattering being the most efficient energy loss mechanism.

We can combine the ratio and transfer fraction to calculate the bulk energy fraction transferred to the halo over the Hubble time. As alluded to previously, the energy is transferred via the secondary particles created during the injected particle journey through the halo. While we here are not in the position to give rigorous treatment to the injection of the secondary particle, we do observe these to be of considerably lower energy than the original ''parent" particle. This holds in particular for the dominant energy transfer process of electrons and positrons undergoing IC scattering and producing lower energy photons. We make the assumption that the secondary particles produced through inverse Compton scattering, transfer their energy more readily to the gas in the halo and subsequently set an energy absorption fraction of $f_{abs} = 0.1$ with the rest of the energy escaping [compare the photon escape fractions of ionising galaxies \citet{Mitra2012}, \citet{Benson2013}], so that we have,
\begin{eqnarray}
F_{eff} = R(M, z)T(M, z)f_{abs}.
\end{eqnarray}
The precise energy absorption fraction from secondary particles, as parameterised by $f_{abs}$ in eq. 16, is critical in determining the impact dark matter annihilation may have on the halo's gas component. Since we don't rigorously treat secondary energy absorption in this work, it is difficult to accurately quantify the impact these particles have on the overall energy deposition into the halo. For example, \citet{Spolyar2008} in their analysis of GeV range neutralino type dark matter, found that in order for there to be efficient energy transfer from the injected particles (including secondaries) notably higher gas densities, such are found after collapse of the proto-stellar core, are required. It may however be worth noting that Spolyar's work does not take into consideration electrons and positrons losing energy through inverse Compton scattering off CMB photons. Instead high energy charged particles solely emit Bremsstrahlung radiation which in turn undergoes electron-positron pair-creation, thus triggering an electromagnetic cascade \citep{PDG2006}. While these cascades also down scatter the injected energy, the process produces secondary particles with high energies, particularly during the early stages of the cascade. When taking secondaries into account, electromagnetic cascades are perhaps not as efficient at energy transfer as the IC scattering mechanism at high redshifts, because the latter produces low energy secondaries more readily. These have interaction cross-sections bigger than those of the EM cascade secondaries, corresponding to a higher energy transfer rate. Inverse Compton scattering can also be thought of as producing an energy spectrum of secondaries comparable to that of particles injected by annihilation of much smaller (keV - MeV) dark matter candidates, which have been found to evacuate notable fractions of gas from small halos \citep{Ripamonti2007}. This further motivates the relatively high absorption fraction used here. We return to discuss different values of $f_{abs}$ in \S6.1.   
 
\begin{figure*}
\centering
\includegraphics[trim= 0 0 0 0, clip, scale = 0.5]{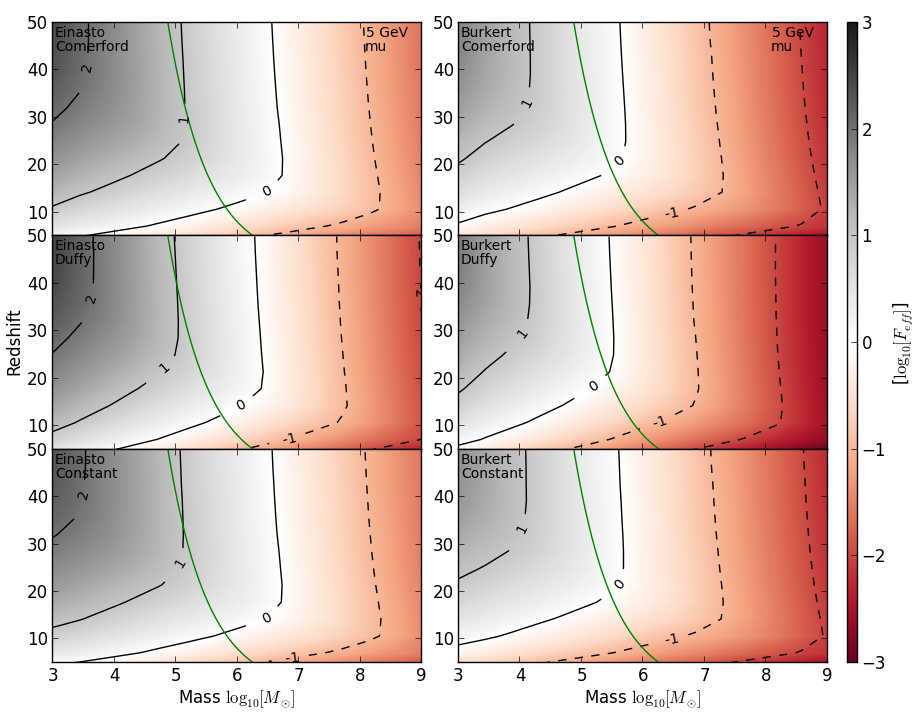}
\caption{Ratio of energy produced by dark matter annihilation over the Hubble time deposited into the halo, to the halo's gravitational binding energy for a 5 GeV dark matter particle annihilating via a muon. The left hand side shows Einasto profiles with Comerford, Duffy and constant mass concentration relations respectively with $f_{abs}=0.1$, while the right shows the same with a Burkert profile. The black contour lines again correspond to the values of the colourbar. The green line is the critical mass for which molecular cooling is possible. Halo models to the right of the line can cool efficiently.} 
\label{Ratios}
\end{figure*}

\begin{figure*}
\centering
\includegraphics[scale = 0.5]{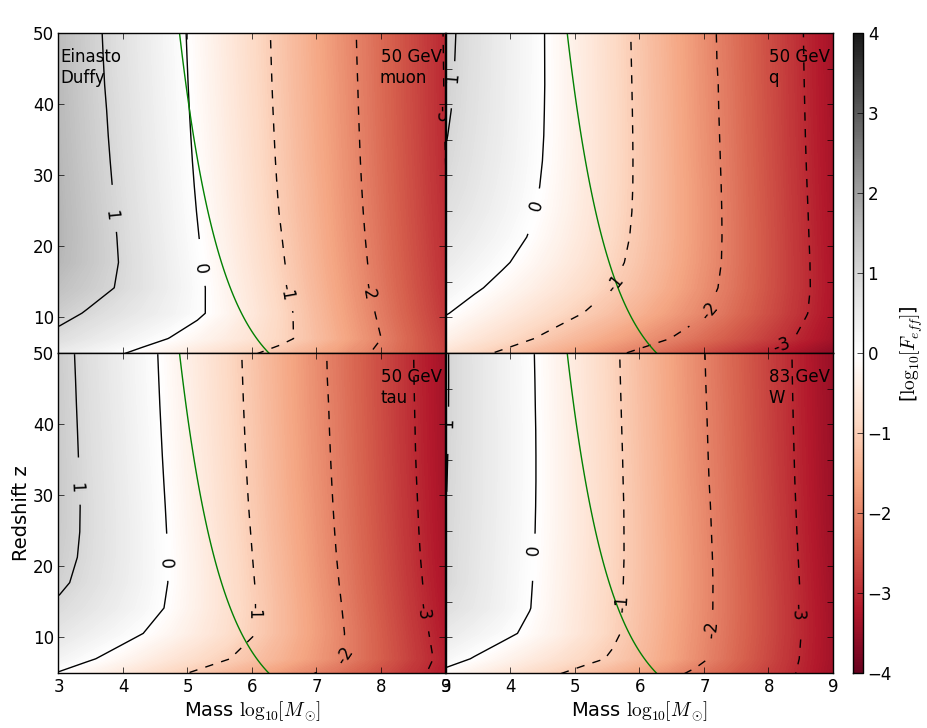}
\caption{Ratio of energy produced by dark matter annihilation over the Hubble time to the halo's gravitational binding energy for a halo with an Einasto profile and Duffy mass concentration relation and $f_{abs}=0.1$. The panels show respectively, a 50 GeV particle annihilating via muon, quark, and tau particles as well as 83 GeV particle annihilation via a W boson. Again the green line is the critical halo mass for molecular cooling, while the black contour lines correspond to the colourbar. Halo models to the right of the line can cool efficiently.} 
\label{Ratios}
\end{figure*}

Figures 5 and 6 show this effective energy transfer fraction, assuming $f_{abs} = 0.1$. In Figure 5 we utilise a 5 GeV dark matter particle annihilating to muon and show $F_{eff}$ for various halo models. The left hand panel shows results for Einasto profiles, with our three mass-concentration models, and the right hand panel shows results for the Burkert model. While the overall behaviour is the same for all halo models, we find the cuspy Einasto model to be more efficient at self-heating. In a similar vein, the mass-concentration relation which produces the highest value for $c$ produces the greatest $F_{eff}$ at fixed redshift and halo mass, indicating that the more concentrated the halo the more efficient is the energy transfer process.

At high redshifts, star formation has not yet disassociated molecular Hydrogen, providing a cooling channel in halos of $10^{5}-10^{6} M_{\odot}$ \citep{Haiman1997}. Also included in Figure 5 is the critical mass above which halos undergo molecular hydrogen cooling (green curve) \citep{Loeb2006}. Halos to the left of the curve do not cool and therefore cannot collapse and form stars. We note that for all models there is a region between $z = 15-50$ and for halos $10^{5}-10^{6}\mathrm{M}_{\odot}$ in which molecular hydrogen cooling is possible but $F_{eff}>1$. This opens the possibility that dark matter annihilation could have a significant impact on the gas chemistry in these systems and by extension on other internal structure formation.

Figure 6 shows similar plots for different annihilation channels and an Einasto profile with a Duffy mass-concentration relation. The tau, muon and quark cases all correspond to 50 GeV dark matter particles while the W boson case shows a 83 GeV particle. In all cases we find that at high redshift, $F_{eff} \sim 1$ either coincides with the molecular cooling line or lies to the left of it, suggesting a smaller impact from larger dark matter particle candidates. At the same time, we note that while star formation may not be impacted in the largest halos for this dark matter model, they nevertheless act as both sources and sinks for ionising radiation. This should be taken into account when including annihilating dark matter in reionisation calculations.

Finally while we find that at high redshift the energy deposition behaviour is comparable between all four models (with the mu path being the most efficient), there is greater variation at low redshift. This is because as the mechanism with which the injected particles lose energy becomes less efficient, their sensitivity to the injected particle's initial energy increases and we see variation in $F_{eff}$ due to the differences in their dark matter model's injected particle energy distributions.

\subsection{Uncertainty due to $f_{abs}$}
We consider the uncertainty in our estimate in the secondary particle energy absorption fraction by comparing the energy depositions plots for a range of $f_{esc}$, shown in figure 7. From top to bottom,  panels show $f_{abs} = 0.001, 0.01$ and $0.1$, in all cases the halo hosts an Einasto profile with a Duffy mass-concentration relation. We note the impact $f_{abs}$ has on the energy transfer into the halo's gas. In particular in the case in which only a $f_{abs} = 0.001$ of the energy carried by secondary particles is transferred to the halo, the impact dark matter annihilation has on heating the halo's gas becomes significantly reduced. This reaffirms the importance of careful future treatment of the secondary particles within the halo. 
\begin{figure}
\centering
\includegraphics[trim= 0.1 0 0 0, clip,scale=0.35]{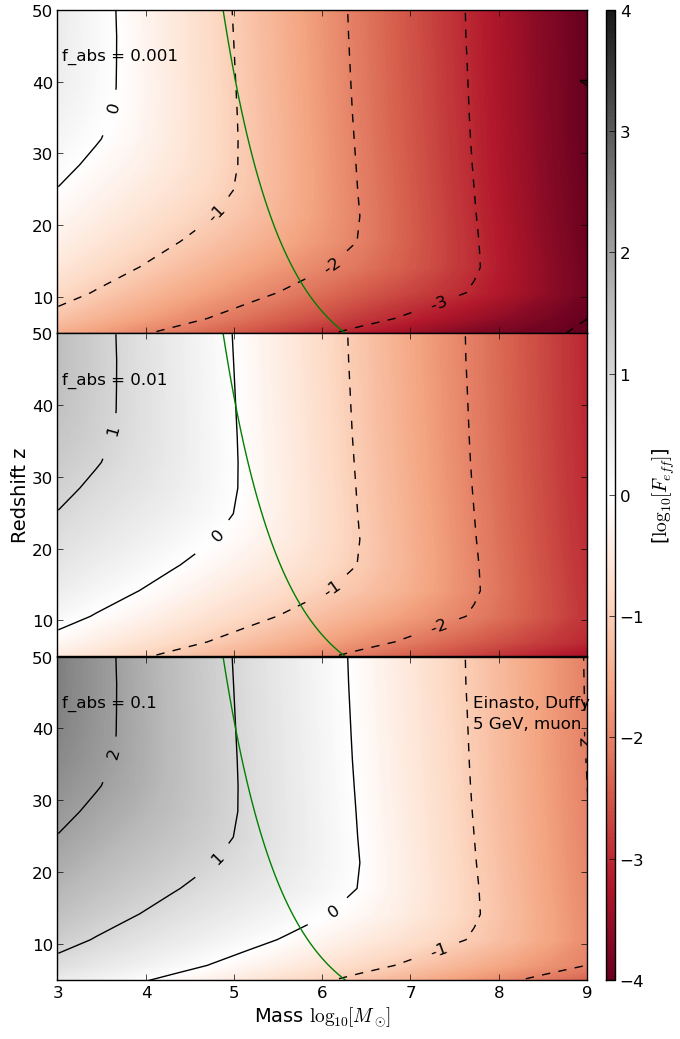}
\caption{Ratio of energy produced by dark matter annihilation over the Hubble time deposited into the halo, to the halo's gravitational binding energy for different secondary particle absorption fractions. From top to bottom, $f_{abs} = 0.001, 0.01, 0.1$ In all cases we are considering a 5 GeV dark matter particle annihilating via the muon channel in a halo with an Einasto profile and Duffy mass concentration relation. Again the contour lines correspond to the plot's colourbar.}
\end{figure}

\subsection{MEDEA}
The above calculation gives an estimate of the gross energy transfer from DM annihilation to gas within the halo. There is a further partition into the energy that is channelled to heating, ionisation and the creation of Lyman alpha photons. While the stopping power and cross-sections used here are averaged quantities that don't track the secondary cascade particles, one can calculate the spectrum of photons produced through the IC process. As this is also the largest channel through which energy is deposited into the halo we can use this in conjunction with the MEDEA2 code to give an indication of the breakdown of the deposited energy. 

\begin{figure*}
\includegraphics[scale=0.55]{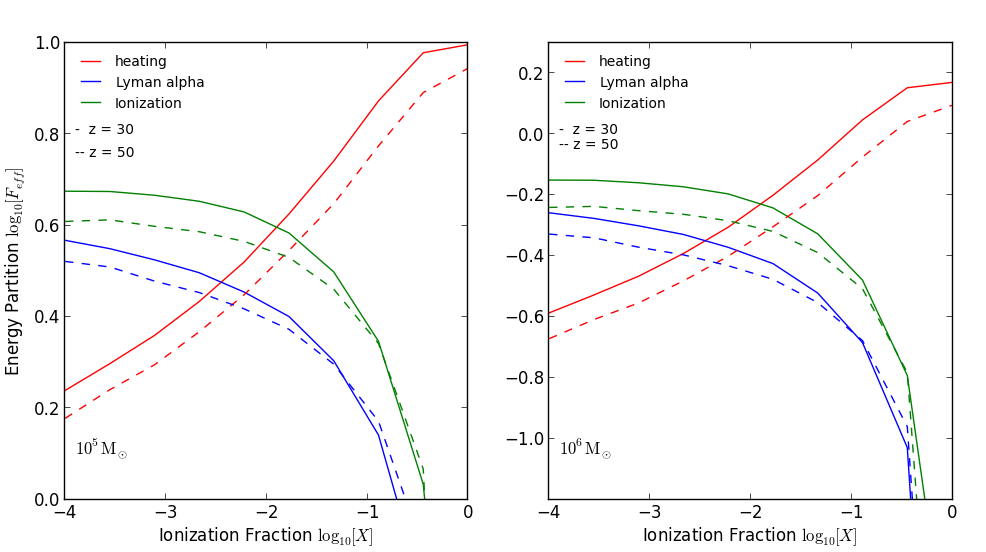}
\caption{Ratio of dark matter annihilation energy and halo binding energy over Hubble time with MEDEA I energy deposition for a $10^5 \mathrm{M}_{\odot}$ (left) and $10^{6} \mathrm{M}_{\odot}$ (right) halo with an Einasto profile and Duffy mass-concentration relation and $f_{abs}=0.1$. Continuous lines indicate z = 30 and the dashed lines z=50. Red indicates energy into heating, blue Lyman alpha photons and green ionisation.}
\end{figure*}

The MEDEA2 code uses a Monte Carlo method to randomly sample physical processes such as interaction probabilities and cross-sections to trace the path of an injected particle through primordial gas (atomic hydrogen and helium) with a homogeneous density distribution. It also tracks secondary particles created during particle cascades. For a particle (electron, positron or photon) with specified energy at a given redshift (gas density), it returns the energy partition of the injected particle into heating, H and He ionisation and Lyman alpha photons. MEDEA2 produces these for a specifiable range of gas-ionisation fractions. 

To account for the secondary particles, we produced MEDEA2 partitions for the energy range of the up-scattered, IC photons. The MEDEA code assumes a uniform gas distribution and is so not representative of the variable density of the halo's gas component. To incorporate this density variation into our calculation, we also produced corresponding partitions at a range of redshifts corresponding to the range of gas densities found in the halo. We found that overall there is little difference between the partition fractions produced for different gas densities as photons of the energy range considered here, are absorbed efficiently by the gas even at low densities, and produce little to no free-streaming particles. We also note that in contrast to our treatment of the primary, injected particles where the gas density is fixed with radius and the CMB radiation varies with redshift, the gas density and CMB photon number densities in MEDEA2, evolve conjointly with redshift due to expansion. The use of MEDEA2 with a redshift density proxy remains valid for the case of the particles under consideration here as low energy photons do not interact with the CMB field. We subsequently used the photon MEDEA partitions to produce energy weighted averages of the partition fractions.

Figure 8 shows how the energy transferred would be partitioned into heating, ionisation and Lyman alpha photons for different ionisation fractions of the halo gas, assuming IC scattering is the dominant component of energy deposition. We compare a $10^{5}\mathrm{M}_{\odot}$ halo on the left with a $10^{6}\mathrm{M}_{\odot}$ halo on the right. In both cases we show a 5 GeV dark matter model annihilating via muons and $f_{abs}=0.1$. The solid lines correspond to redshift 30 and the dashed to redshift 50. 

\section{Discussion}
Using a simple analytic treatment, we have calculated the degree to which the small DM halos at high redshift which are thought to host the first stars heat and ionise themselves through the annihilation of dark matter. We find that the total energy produced by the annihilation process over the Hubbble time exceeds the gravitational binding energy of the halo's gas for structures with mass less than $\sim 10^{8}\mathrm{M}_{\odot}$. However, the high-energy stable particles produced in this process do not readily interact with the surrounding gas so energy is transferred through secondary particles.  For a secondary particle energy absorption fraction of $f_{abs}=0.1$ and taking the critical mass for molecular cooling into consideration, we find that there is a parameter space in which primordial gas inside a $10^{5}$-$10^{6}\mathrm{M}_{\odot}$ halo above redshift 20 could cool but where the injected energy from dark matter still exceeds the binding energy of the gas. This could lead to the disruption of early star formation, and a delay in the formation of the first galaxies.

We find that the lighter dark matter particle masses with the muon annihilation channel are the most effective inhibitors of star formation. Concentrated halos that display some sort of cusp like behaviour are also more efficient at self-heating due to the square dependence of the dark matter power on the density distribution.  

This complements the work of \citet{Ripamonti2007}, in which the authors found that dark matter annihilation/ decay may lead to a substantial evacuation of gas from halos with mass less than $10^{6}\mathrm{M}_{\odot}$. While their work implemented lighter dark matter candidates with annihilation/decay products that generally interact with the gas in the halo more readily, our treatment of the injected relativistic particles following \citet{Evoli2012} shows that disruption of the star formation process maybe still be a possibility, even for heavier dark matter models. 

The efficiency with which the injected particles transfer their energy to the halo is of key importance in this calculation. However, the complete description of the energy loss of not only the primary injected particles but also their secondary progeny is beyond the scope of this work. In particular, the energy absorption fraction, $f_{abs}$, of the secondary particles plays a crucial role in determining the overall impact dark matter annihilation has on heating the halo's gas. The full realisation of the energy transfer process within the bounds of the halo, including non-static gas conditions, will be addressed in a future paper. Similarly, a detailed description of the impact of the extra injected energy on the halo's gas chemistry and what consequences this would have for structure formation are also left for future work. 

\section*{Acknowledgments}
We wish to thank Carmelo Evoli for access and assistance with the MEDEA code. We also thank the anonymous referee for their suggestions and input. KJM is supported by the Australian Research Council Discovery Early Career Research Award.

\bibliographystyle{mn2e}	
\bibliography{halos}

\appendix
\section{Detailed PYTHIA Outputs}
In Figure A1 we show detailed final energy distributions from PYTHIA for electrons, positrons, and photons for the different annihilation channels. Each plot shows the fraction of the centre of mass energy of the annihilation process carried by particles with different energy averaged over 100000 events.

\begin{figure}
\centering
\includegraphics[scale=0.6]{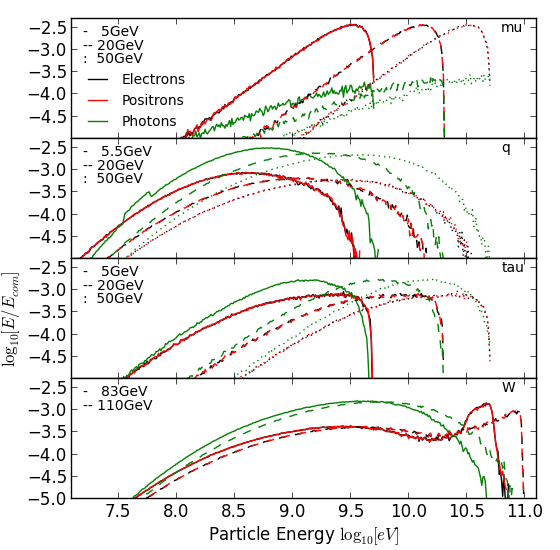}
\caption{Averaged fraction of the center of mass energy carried by photons (green), electrons (black) and positrons (red) of energy $E$ for 100000 neutralino annihilation events in PYTHIA. The upper three panels correspond from top to bottom to annihilation to mu, quark, and tau particles where in each, the solid line gives the final, stable particle distribution for a 5GeV (or 5.5GeV in case of the quark) dark matter particle, the dashed a 20GeV particle and the dotted the 50GeV case. The lowest panel shows annihilation to a W boson where here the solid line is the 83GeV model and the dashed the 110 GeV model.}
\label{Electrons}
\end{figure}

\section{Energy Transfer Details}
We here illustrate in greater detail the mechanisms \citep{Evoli2012} via which the injected particle transfer energy to the halo.

The dark matter model considered in this work annihilate predominantly to high energy particles. While the precise energy spectral density is dependent on the annihilation channel and the dark matter particle mass (see figure A1), electrons, positrons and photons are generally injected with energy in the order of GeV and above. 
\subsection{Photon}
The mechanisms via which photons lose energy to the gas are photo-ionisation, Compton scattering and electron-positron pair-creation. For photons of the energy range discussed here, pair-creation is the dominant interaction.
Since interaction cross-sections are larger for low energy particles, the energy transfer rate of photons is driven foremost by the dense gas at the core of the halo and overall the contribution from injected photons to the total energy deposited is secondary to that from electrons and positrons. 

\subsection{Electrons and Positrons}
Electrons and positrons lose energy through interaction with the halo's gas component, as well as the CMB photon background. In the former case, this is enabled through Bremsstrahlung and collisional interactions (ionisation and excitation). Loss through Bremsstrahlung is the dominant process for high energy particles and collisional interaction for low energy particles, with the critical energy $E_{crit}\sim340 \mathrm{MeV}$ giving the cross-over between the two regimes. We note that for the injected electrons, Bremsstrahlung is thus the more prevalent process of the two. Energy loss is more efficient in high density regions such as the halo's core and is thus sensitive to the baryonic profile. 

Interaction with the CMB photon background occurs through inverse Compton (IC) scattering, in which some of the electron's energy is transferred to the scattering photon. The energy transferred is dependent on both the original energy of both the photon and electron, with $\frac{ \langle \nu_{f} \rangle }{\nu_{i}} = \frac{4}{3} \gamma^{2}$ where $\langle \nu_{f} \rangle, \nu_{i}$ are the average energy of the up-scattered photon and energy of the original photon and $\gamma$ the boost factor of the electron. The electron will undergo multiple IC scattering events. The efficiency of this mechanism will diminish with redshift due to the reduction of CMB photon number density, as well as the drop in the CMB temperature. At high redshift this is the dominant electron energy loss mechanism. This behaviour can be observed in the figure 3, specifically the efficient energy loss for high energy electrons at high redshift, independent of the baryonic profile. 

While IC scattering leads to loss of energy for the electron, this energy is not directly transferred to the halo. Instead the secondary, up-scattered photons created in the process interact with the gas and deposit energy in the form of heat, photo-ionisation and Lyman alpha photons This mechanism is far more efficient than the energy transfer from initially injected particles due to the up-scattered photon's low energy (up to UV/soft X-ray). Our calculation of the energy transfer does not extend to a precise treatment of these secondary particles. We refer to the discussions of escape fractions of photons from early star forming halos as well as the results produced by MEDEA2 to motivate our estimate of the secondary photon energy transfer rate of $f_{abs} = 0.1$, see \S 6. We will in future work determine the distribution of energy within the halo.

\section{Baryonic Profile Comparison}
\begin{figure}
\centering
\includegraphics[trim= 0 0 0 0, clip,scale=0.6]{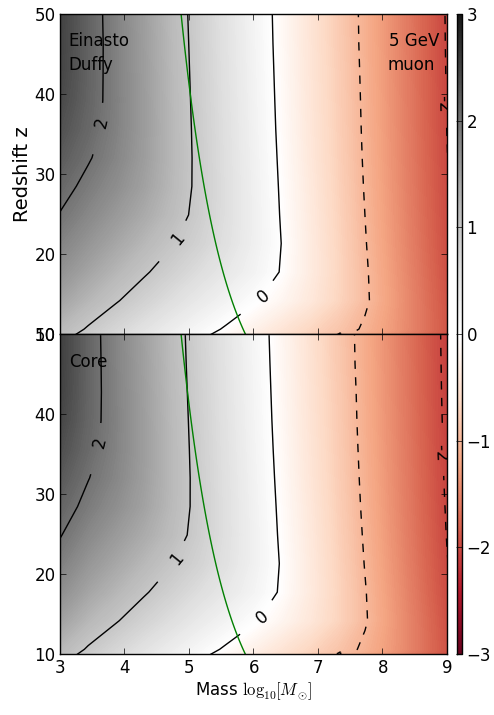}
\caption{Ratio of energy produced by dark matter annihilation over the Hubble time deposited into the halo, to the halo's gravitational binding energy for a 5 GeV dark matter particle annihilating via a muon. The upper panel shows  and Einasto-like baryonic profile with a Duffy mass-concentration relation and $f_{abs}=0.1$, while the lower shows the same with a baryonic core profile.}
\end{figure}
In figure C1, we show the comparison between the ratio of energy produced by dark matter annihilation over the Hubble time and transferred into the halo, and the gravitational binding energy, for a case in which the baryonic profile traces that of the dark matter distribution (upper panel) and one for which the baryonic component forms a core (lower panel). The two models produce comparable results, with the baryonic core model being only marginally less efficient, as expected. This occurs because the IC scattering mechanism is not heavily reliant on the dense gas region at the centre of the halo to effect energy transfer. 

\end{document}